\documentclass[pra,aps,twocolumn,showpacs]{revtex4}
\usepackage{times,epsfig,amssymb,amsfonts,amsmath,graphicx}

\newcommand{\ket}[1]{|#1\rangle}

\newcommand{\beq}{\begin{equation}}
\newcommand{\eeq}{\end{equation}}
\newcommand{\beqa}{\begin{eqnarray}}
\newcommand{\eeqa}{\end{eqnarray}}

\begin{document}

\title{Polygamy relation for the R\'{e}nyi-$\alpha$ entanglement of assistance in multi-qubit systems}

\author{Wei Song$^{1}$}
\author{Jian Zhou$^{2}$}
\email{jianzhou8627@163.com(Corresponding~author)}
\author{Ming Yang$^{3}$}
\author{Jun-Long Zhao$^{3}$}
\author{Da-Chuang Li$^{1}$}
\author{Li-Hua Zhang$^{4}$}
\author{Zhuo-Liang Cao$^{1}$}

\affiliation{$^{1}$ Institute for Quantum Control and Quantum Information, and School
of Electronic and Information Engineering, Hefei Normal University, Hefei 230601, China\\
$^{2}$Department of Electronic Communication Engineering, Anhui Xinhua University, Hefei 230088,
People¡¯s Republic of China\\
$^{3}$School of Physics and Material Science, Anhui University, Hefei 230601, China\\
$^{4}$School of Physics and Electrical Engineering, Anqing Normal University, Anqing 246133, China\\}
\begin{abstract}
We prove a new polygamy relation of multi-party quantum entanglement in terms of R\'{e}nyi-$\alpha$ entanglement of assistance for $\left( {\sqrt 7 - 1} \right)/2\leq\alpha  \leq \left( {\sqrt 13  - 1} \right)/2$. This class of polygamy inequality reduces to the polygamy inequality based on entanglement of assistance since R\'{e}nyi-$\alpha$ entanglement is a generalization of entanglement of formation. We further show that the polygamy inequality also holds for the $\mu$th power of R\'{e}nyi-$\alpha$ entanglement of assistance.
\end{abstract}


\maketitle

One fundamental property of quantum entanglement is in its limited shareability in multi-party
quantum systems\cite{hor09rmp}. For example, if the two subsystems are more entangled with each other, then they will share a less amount of entanglement with the other subsystems with specific entanglement measures. This restricted shareability of entanglement is named as the monogamy of entanglement (MoE). The concept of monogamy is an essential feature allowing
for security in quantum key distribution\cite{Pawlowski10pra}. It also
plays an important role in many field of physics such
as foundations of quantum mechanics\cite{Bennett14,Toner09,Seevinck10qip}, condensed matter
physics\cite{Ma11np,Saez13prb}, statistical mechanics\cite{Bennett14}, and
even black-hole physics\cite{Susskind13,Lloyd14}. Monogamy inequality was first built for three-qubit
systems using tangle as the bipartite entanglement measure\cite{ckw00pra}, and generalized into multi-qubit systems in terms of various entanglement measures\cite{osb06prl,off08pra,eos09pra,raj10pra,Cornelio13,Regula14,byw09pra,Chi08,Yu08,Osterloh15,Eltschka15,ade06njp,hir07prl,ade07prl,koa04pra,fan07pra,kim09pra,he15pra,choi15pra,yluo15ap,kim10jpa,cor10pra,loh06prl,oli14pra,bxw14prl,zhu14pra,bxw14pra,gao15sr,Song16pra,Yuan16sr,Luo16pra,Tian16sr}.

On the other hand, the assisted entanglement, which is a dual concept to bipartite
entanglement measures, is known to have a dually monogamous or polygamous property in multiparty quantum systems. The polygamous property can be regarded as another kind of entanglement constraints in multi-qubit systems, and Gour \emph{et al}\cite{Gour05pra} established the first dual monogamy inequality or polygamy inequality for multi-qubit systems using concurrence
of assistance (CoA). For a three-qubit pure state ${\left| \psi  \right\rangle _{A_1 A_2 A_3 } }$, a polygamy inequality was introduced as:

\beqa\label{q1}
C^2 \left( {\left| \psi  \right\rangle _{A_1 |A_2 A_3 } } \right) \le \left[ {C^a \left( {\rho _{A_1 A_2 } } \right)} \right]^2  + \left[ {C^a \left( {\rho _{A_1 A_3 } } \right)} \right]^2,
\eeqa
where CoA for a bipartite state ${\rho _{AB} }$ is defined as: $C^a \left( {\rho _{AB} } \right) = \max \sum\nolimits_i {p_i C\left( {\left| {\psi _i } \right\rangle _{AB} } \right)}$, with the maximum is taken over all possible pure state decompositions of ${\rho _{AB} }= \sum\nolimits_i {p_i \left| {\psi _i } \right\rangle _{AB} \left\langle {\psi _i } \right|}$ and ${C\left( {\left| {\psi _i } \right\rangle _{AB} } \right)}$ denotes the concurrence\cite{woo98prl} of ${\left| {\psi _i } \right\rangle _{AB} }$. Furthermore, it is shown that for any pure state $\left| \psi  \right\rangle _{A_1 A_2  \cdots A_n }$ in a $n$-qubit system\cite{Gour07jmp}, we have

\beqa\label{q2}
C^2 \left( {\left| \psi  \right\rangle _{A_1 |A_2  \cdots A_n } } \right) \le \left[ {C^a \left( {\rho _{A_1 A_2 } } \right)} \right]^2  +  \cdots  + \left[ {C^a \left( {\rho _{A_1 A_n } } \right)} \right]^2.\nonumber\\
\eeqa

Later, polygamy inequalities was generalized in terms of Tsallis entanglement of assistance(TEoA)\cite{Kim10pra1} or unified entanglement of assistance\cite{Kim12pra}, and polygamy inequalities in higher-dimensional systems were also shown using the entanglement of assistance(EoA)\cite{Buscemi09pra} or TEoA\cite{Kim16pra}. In this paper, we establish a new polygamy relation of multi-party quantum entanglement in terms of R\'{e}nyi-$\alpha$ entropy (ER$\alpha$E)\cite{kim10jpa}. As an important generalization of entanglement of formation(EoF), ER$\alpha$E is a well-defined entanglement measure which has a continuous spectrum parametrized by the non-negative real parameter $\alpha$. It reduces to the standard EoF when $\alpha$ tends to $1$. Thus our polygamy inequalities including previous polygamy relation of EoF as a special case\cite{Buscemi09pra}. Furthermore, we generalize the polygamy inequalities in terms of the $\mu$th power of R\'{e}nyi-$\alpha$ entanglement of assistance.

For a bipartite pure state $\left| \psi  \right\rangle _{AB}$, the ER$\alpha$E is defined as

\beq\label{q3}
E_{\alpha}(\left| \psi  \right\rangle _{AB}):= S_\alpha(\rho_A) := \frac{1}{1-\alpha}\log (\mbox{tr}\rho _A^\alpha),
\eeq
where $S_\alpha(\rho_A)$ is the R\'{e}nyi-$\alpha$ entropy. The R\'{e}nyi-$\alpha$ entropy has found important applications in characterizing quantum phases with
differing computational power \cite{Cui12nc}, ground state properties
in many-body systems \cite{Franchini14prx}, and topologically ordered states
\cite{Flammia09prl,Halasz13prl}. The ER$\alpha$E of a bipartite mixed state $\rho _{AB}$ can be defined using the convex roof technique
\beqa\label{q4}
E_\alpha(\rho _{AB})=\mbox{min} \sum_i p_i E_\alpha(\ket{\psi _i }_{AB}).
\eeqa
It is known that R\'{e}nyi-$\alpha$ entropy converges to the von Neumann entropy when $\alpha$ tends to $1$. So the entanglement R\'{e}nyi-$\alpha$ entropy reduces to the EoF when $\alpha$ tends to $1$. For any
two-qubit state ${\rho _{AB} }$ with $\alpha  \ge \left( {\sqrt 7  - 1} \right)/2$, there exist an analytic formula of ER$\alpha$E\cite{kim10jpa,yxwang16pra}

\beqa\label{q5}
E_\alpha  \left( {\rho _{AB} } \right) = f_\alpha  \left( {C\left( {\rho _{AB} } \right)} \right),
\eeqa
where
\beqa\label{q6}
f_\alpha  \left( x \right) = \frac{1}{{1 - \alpha }}\log \left[ {\left( {\frac{{\Theta \left( x \right)}}{2}} \right)^\alpha   + \left( {\frac{{\Xi \left( x \right)}}{2}} \right)^\alpha  } \right],
\eeqa
with $\Theta \left( x \right) = 1 + \sqrt {1 - x^2 },\Xi \left( x \right) = 1 - \sqrt {1 - x^2 } $.

As a dual concept to ER$\alpha$E, we define the R\'{e}nyi-$\alpha$ entanglement of assistance(REoA) as

\beqa\label{q7}
E_\alpha ^a \left( {\rho _{AB} } \right): = \max \sum\nolimits_i {p_i } E_\alpha  \left( {\left| {\psi _i } \right\rangle _{AB} } \right),
\eeqa
where the maximum is taken over all possible pure state decompositions of ${\rho _{AB} }= \sum\nolimits_i {p_i \left| {\psi _i } \right\rangle _{AB} \left\langle {\psi _i } \right|}$.

For $0<\alpha<1$, we can derive a upper bound of REoA. From the definition of entanglement of REoA, we have

\beqa\label{q7}
 E_\alpha ^a \left( {\rho _{AB} } \right) &=& \max \sum\nolimits_i {p_i E_\alpha  \left( {\left| {\psi _i } \right\rangle _{AB} } \right)}  \nonumber\\
  &=& \max \sum\nolimits_i {p_i S_\alpha  \left( {\rho _{iA} } \right)}  \nonumber\\
  &\le& S_\alpha  \left( {\sum\nolimits_i {p_i \rho _{iA} } } \right) \nonumber\\
  &=& S_\alpha  \left( {\rho _A } \right),
\eeqa
where ${\rho _{iA} }$ is the reduced density matrix of ${\left| {\psi _i } \right\rangle _{AB} }$, and the inequality holds due to the concave property of $S_\alpha  \left( \rho  \right)$ for $0<\alpha<1$\cite{Bassat78,Mosonyi11,Debarba17}. Similarly, we can derive $E_\alpha ^a \left( {\rho _{AB} } \right)\le S_\alpha  \left( {\rho _B } \right)$. Thus we have $E_\alpha ^a \left( {\rho _{AB} } \right)\le \min \{ S_\alpha  \left( {\rho _A } \right),S_\alpha  \left( {\rho _B } \right)\}$

Before showing the main result of this paper, we first
give two lemmas as follows.

\emph{Lemma 1.} For any two-qubit state $\rho _{AB}$ and $\alpha  \ge \left( {\sqrt 7  - 1} \right)/2$, we have
\beqa\label{q8}
E_\alpha^a \left( {\rho _{AB} } \right) \ge f_{\alpha}\left( {C^a \left( {\rho _{AB} } \right)} \right),
\eeqa
where $E_\alpha^a \left( {\rho _{AB} } \right)$ and ${C^a \left( {\rho _{AB} } \right)}$ are the REoA and CoA of ${\rho _{AB} }$, respectively.

\emph{Proof.} Suppose that the optimal decomposition for ${C^a \left( {\rho _{AB} } \right)}$ is $\left\{ {p_{i,} \left| {\psi _i } \right\rangle _{AB} } \right\}$, we have
\beqa\label{q9}
 f_{\alpha}\left( {C^a \left( {\rho _{AB} } \right)} \right) &=& f_{\alpha}\left( {\sum\nolimits_i {p_i C\left( {\left| {\psi _i } \right\rangle _{AB} } \right)} } \right) \nonumber\\
  &\le& \sum\nolimits_i {p_i f_{\alpha}\left( {C\left( {\left| {\psi _i } \right\rangle _{AB} } \right)} \right)}  \nonumber\\
  &=& \sum\nolimits_i {p_i E_\alpha \left( {\left| {\psi _i } \right\rangle _{AB} } \right)}  \le E_\alpha^a \left( {\rho _{AB} } \right),
\eeqa
where in the first inequality we have used the convex property of $f_{\alpha}(x)$ as a function of $x$ for $\alpha \ge \left( {\sqrt 7  - 1} \right)/2$, and the second inequality is due to the definition of EoA. $\hfill\square$

\begin{figure}
\centering
\includegraphics[scale=0.4,angle=0]{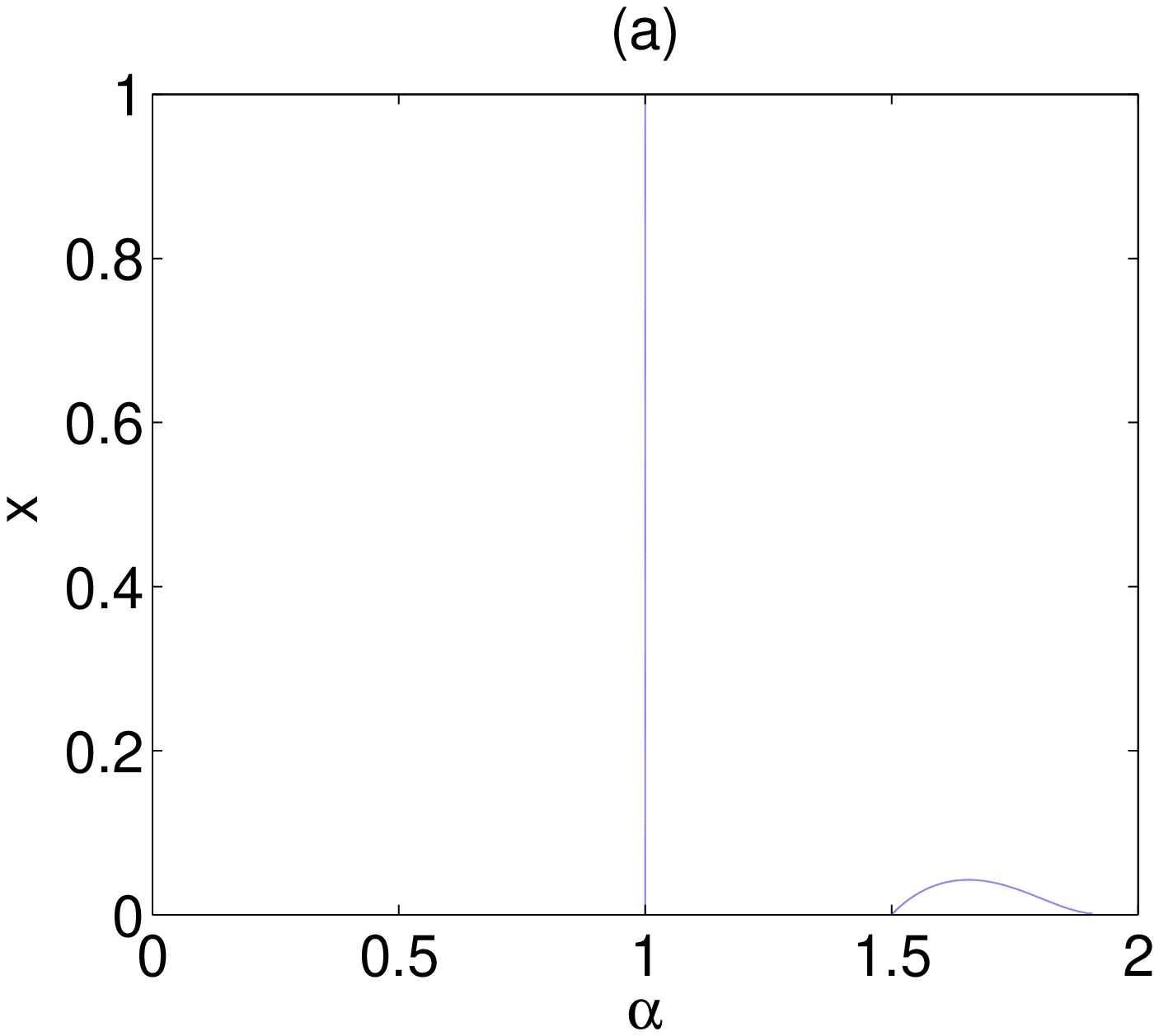}
\includegraphics[scale=0.4,angle=0]{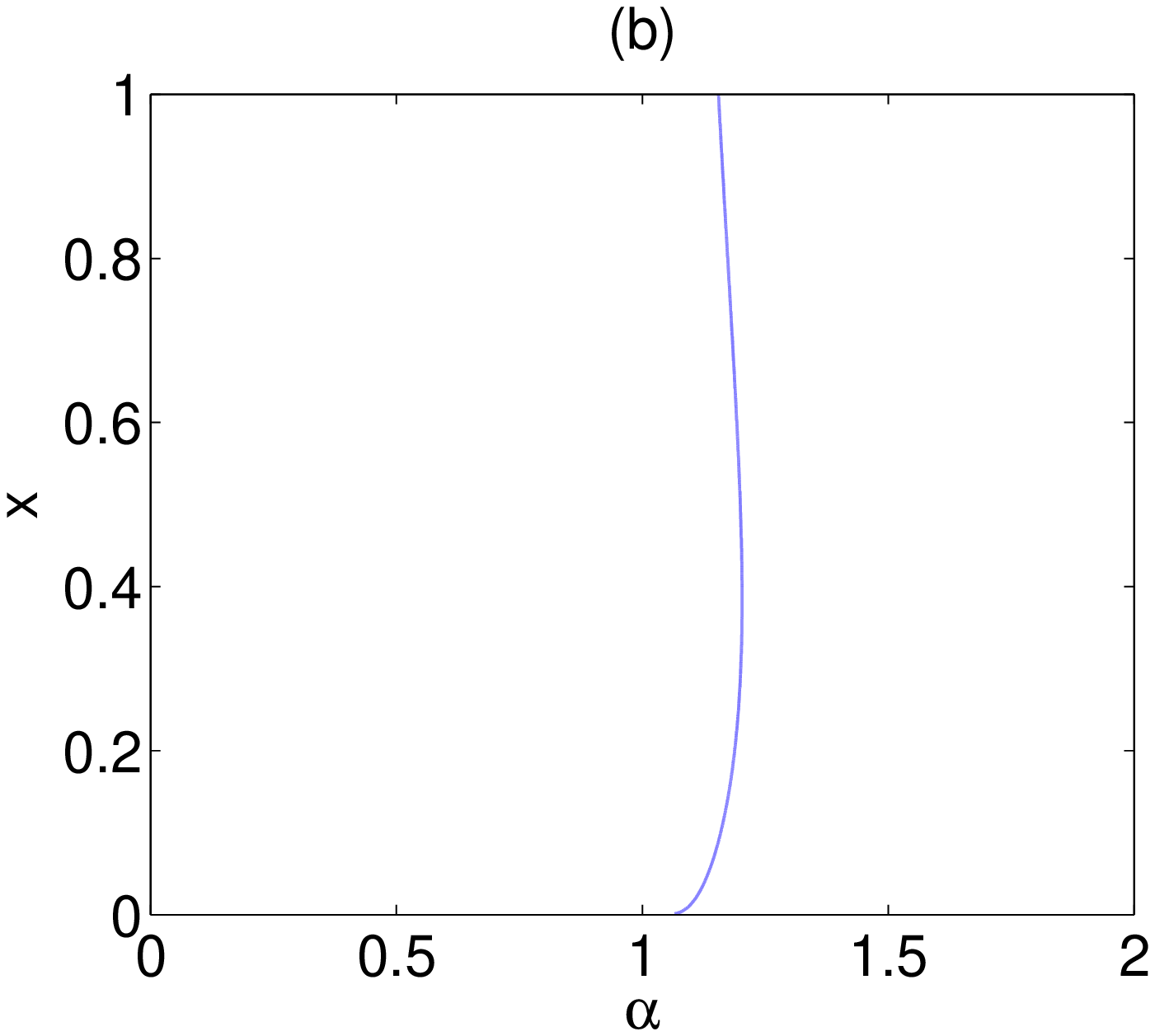}
\caption{(color online) The plots of the equations(a)$\partial h_\alpha  /\partial x =0$; (b)$\partial h_\alpha  /\partial \alpha =0$ for $0 \leq x \leq 1, 0 \leq \alpha \leq 2$.}
\label{f1}
\end{figure}

\emph{Lemma 2.} For any $\left( {\sqrt 7  - 1} \right)/2 \le \alpha  \le \left( {\sqrt {13}  - 1} \right)/2$ and the function $f_\alpha  \left( x \right)$ defined on the domain $\mathcal{D} = \left\{ {\left( {x, y} \right)|0 \leq x,y \leq 1, 0 \leq x^2+y^2 \leq 1} \right\}$, we have
\beqa\label{q10}
f_\alpha  (\sqrt {x^2  + y^2 } ) \le f_\alpha  (x) + f_\alpha  (y).
\eeqa

\emph{Proof.} We define a two-vairable function
\beqa\label{q11}
g_\alpha  \left( {x,y} \right) = f_\alpha  (\sqrt {x^2  + y^2 } ) - f_\alpha  (x) - f_\alpha  (y),
\eeqa
on the domain $\mathcal{D} $. Then it is sufficient to show that $g_\alpha  \left( {x,y} \right)$ is a non-negative function on $\mathcal{D}$. Since $g_\alpha  \left( {x,y} \right)$ is analytic
in the interior of $\mathcal{D}$, and continuous on $\mathcal{D}$, its maximum or
minimum value arises only on the critical points or
on the boundary of $\mathcal{D}$. The critical points of $g_\alpha  \left( {x,y} \right)$
satisfy the condition

\beqa\label{q12}
\nabla g_\alpha  \left( {x,y} \right) = (\frac{{\partial g_\alpha  \left( {x,y} \right)}}{{\partial x}},\frac{{\partial g_\alpha  \left( {x,y} \right)}}{{\partial y}}) = (0,0),
\eeqa
where
\beqa\label{q13}
 &&\frac{{\partial g_\alpha  \left( {x,y} \right)}}{{\partial x}} \nonumber\\
  &=& \frac{{Cx\left[ {\left( {\Theta \left( {\sqrt {x^2  + y^2 } } \right)} \right)^{\alpha  - 1}  - \left( {\Xi \left( {\sqrt {x^2  + y^2 } } \right)} \right)^{\alpha  - 1} } \right]}}{{\sqrt {1 - x^2  - y^2 } \left[ {\left( {\Xi \left( {\sqrt {x^2  + y^2 } } \right)} \right)^\alpha   + \left( {\Theta \left( {\sqrt {x^2  + y^2 } } \right)} \right)^\alpha  } \right]}} \nonumber\\
  &-& \frac{{Cx\left[ {\left( {\Theta \left( x \right)} \right)^{\alpha  - 1}  - \left( {\Xi \left( x \right)} \right)^{\alpha  - 1} } \right]}}{{\sqrt {1 - x^2 } \left[ {\left( {\Xi \left( x \right)} \right)^\alpha   + \left( {\Theta \left( x \right)} \right)^\alpha  } \right]}},
\eeqa
and
\beqa\label{q14}
 &&\frac{{\partial g_\alpha  \left( {x,y} \right)}}{{\partial y}} \nonumber\\
  &=& \frac{{Cy\left[ {\left( {\Theta \left( {\sqrt {x^2  + y^2 } } \right)} \right)^{\alpha  - 1}  - \left( {\Xi \left( {\sqrt {x^2  + y^2 } } \right)} \right)^{\alpha  - 1} } \right]}}{{\sqrt {1 - x^2  - y^2 } \left[ {\left( {\Xi \left( {\sqrt {x^2  + y^2 } } \right)} \right)^\alpha   + \left( {\Theta \left( {\sqrt {x^2  + y^2 } } \right)} \right)^\alpha  } \right]}} \nonumber\\
  &-& \frac{{Cy\left[ {\left( {\Theta \left( y \right)} \right)^{\alpha  - 1}  - \left( {\Xi \left( y \right)} \right)^{\alpha  - 1} } \right]}}{{\sqrt {1 - y^2 } \left[ {\left( {\Xi \left( y \right)} \right)^\alpha   + \left( {\Theta \left( y \right)} \right)^\alpha  } \right]}}.
\eeqa

Suppose that there exists $\left( {x_0 ,y_0 } \right)$ in the interior of $\mathcal{D}$ such that $\nabla g_\alpha  \left( {x_0 ,y_0 } \right) = (0,0)$. From Eq.(\ref{q13}) and Eq.(\ref{q14}), we have

\beqa\label{q15}
l_\alpha  \left( {x_0 } \right) = l_\alpha  \left( {y_0 } \right),
\eeqa
where $l_\alpha  \left( {x } \right)$ is defined as
\beqa\label{q16}
l_\alpha  \left( x \right): = \frac{{\left[ {\left( {\Theta \left( x \right)} \right)^{\alpha  - 1}  - \left( {\Xi \left( x \right)} \right)^{\alpha  - 1} } \right]}}{{\sqrt {1 - x^2 } \left[ {\left( {\Xi \left( x \right)} \right)^\alpha   + \left( {\Theta \left( x \right)} \right)^\alpha  } \right]}},
\eeqa
for $0 < x < 1$. We divide the proof into two cases. We first show that $l_\alpha  \left( {x } \right)$ is a strictly monotone-decreasing function for $0 < x < 1, 1 < \alpha  < \left( {\sqrt {13}  - 1} \right)/2$, then it is sufficient to consider the first-order derivative of $l_\alpha  \left( {x } \right)$. After a direct calculation, we have

\beqa\label{q17}
 \frac{{dl_\alpha  \left( x \right)}}{{dx}} &=& \frac{{\alpha x\left[ {\left( {\Theta \left( x \right)} \right)^{\alpha  - 1}  - \left( {\Xi \left( x \right)} \right)^{\alpha  - 1} } \right]^2 }}{{(1 - x^2 )\left[ {\left( {\Xi \left( x \right)} \right)^\alpha   + \left( {\Theta \left( x \right)} \right)^\alpha  } \right]}} \nonumber\\
  &-& \frac{{(\alpha  - 1)x\left[ {\left( {\Theta \left( x \right)} \right)^{\alpha  - 2}  + \left( {\Xi \left( x \right)} \right)^{\alpha  - 2} } \right]}}{{(1 - x^2 )\left[ {\left( {\Xi \left( x \right)} \right)^\alpha   + \left( {\Theta \left( x \right)} \right)^\alpha  } \right]}} \nonumber\\
  &+& \frac{{x\left[ {\left( {\Theta \left( x \right)} \right)^{\alpha  - 1}  - \left( {\Xi \left( x \right)} \right)^{\alpha  - 1} } \right]}}{{\sqrt {(1 - x^2 )^3 } \left[ {\left( {\Xi \left( x \right)} \right)^\alpha   + \left( {\Theta \left( x \right)} \right)^\alpha  } \right]}}.
\eeqa

\begin{figure}[ptb]
\includegraphics[scale=0.7,angle=0]{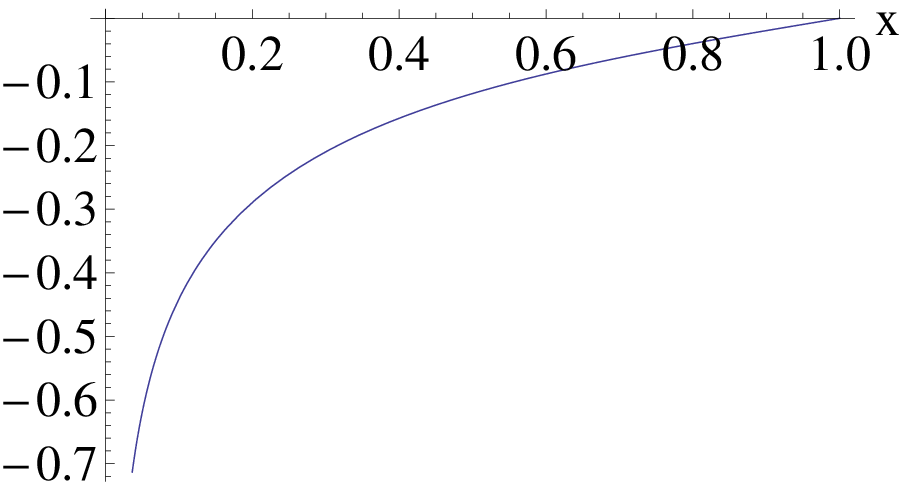}\caption{(color online). The plot of $h_\alpha  \left( x \right)|_{\alpha  = \left( {\sqrt {13}  - 1} \right)/2}$ as a function of $x$ for $0\leq x\leq 1$.}%
\label{fig2}
\end{figure}

\begin{figure}[ptb]
\includegraphics[scale=0.7,angle=0]{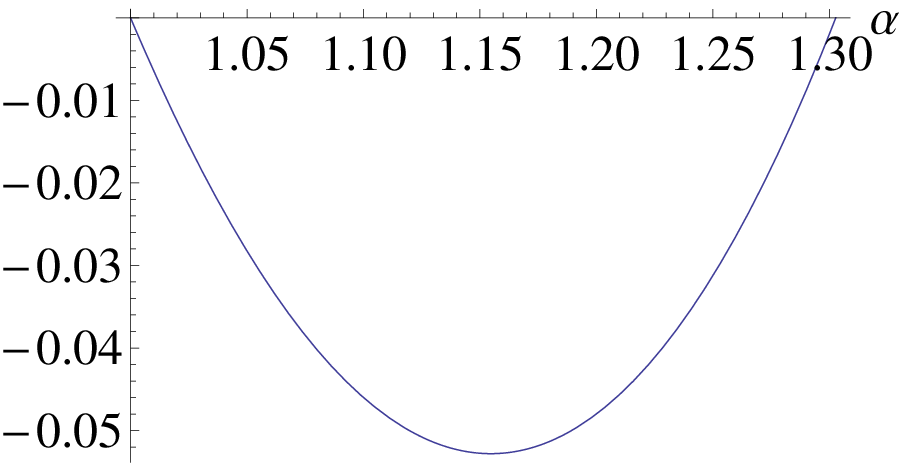}\caption{(color online). The plot of $h_\alpha  \left( x \right)|_{x \to 1}$ as a function of $\alpha$ for $1\leq \alpha \leq \left( {\sqrt {13}  - 1} \right)/2 $.}%
\label{fig3}
\end{figure}

\begin{figure}[ptb]
\includegraphics[scale=0.7,angle=0]{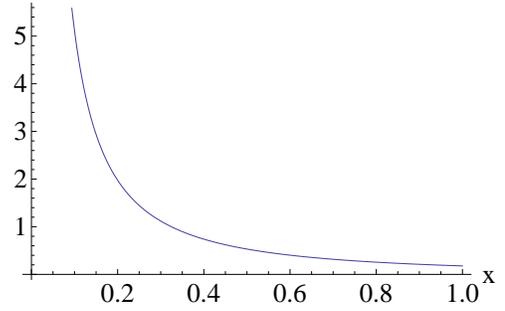}\caption{(color online). The plot of $h_\alpha  \left( x \right)|_{\alpha  = \left( {\sqrt {7}  - 1} \right)/2}$ as a function of $x$ for $0\leq x\leq 1$.}%
\label{fig4}
\end{figure}

In order to show the negativity of the first-order derivative of $l_\alpha \left( {x } \right)$, let us consider the value of the two-variable function $h_\alpha \left( {x } \right):={dl_\alpha  \left( x \right)/dx}$ on the domain $\mathcal{D}_1  = \left\{ {\left( {\alpha ,x} \right)|1 \leq \alpha  \leq \left( {\sqrt {13}  - 1} \right)/2,0 \le x \le 1} \right\}$. The maximum or minimum values of $h_\alpha \left( {x } \right)$ can arise only at the critical points or on the boundary of $\mathcal{D}_1$. The critical points of $h_\alpha \left( {x } \right)$ satisfy the condition $\nabla h_\alpha  \left( x \right) = (\partial h_\alpha  \left( x \right)/\partial \alpha ,\partial h_\alpha  \left( x \right)/\partial x) = (0,0)$. It is shown in Fig.1(a) and (b) that there are no common solutions
on the interior of domain $\mathcal{D}_1$ which indicate that $h_\alpha \left( {x } \right)$ has no critical points on the interior of $\mathcal{D}_1$. Then we consider the function value of $h_\alpha \left( {x } \right)$ on the boundary of $\mathcal{D}_1$. If $\alpha = 1$, we have $h_\alpha  \left( x \right)|_{\alpha  = 1}  = 0$. If $\alpha = \left( {\sqrt {13}  - 1} \right)/2$, we  plot $h_\alpha  \left( x \right)|_{\alpha  = \left( {\sqrt {13}  - 1} \right)/2}$ as a function of $x$ in Fig.2, which illustrates that $h_\alpha  \left( x \right)|_{\alpha  = \left( {\sqrt {13}  - 1} \right)/2}$ is a monotone-increasing function for $0\leq x\leq1$ and obtain its maximum value
$0$ on $x=1$. When $x\to 1$, we have $h_\alpha  \left( x \right)|_{x \to 1}  = 2\left( {q^3  - 4q + 3} \right)/3$ which is always negative for $1 < \alpha  < \left( {\sqrt {13}  - 1} \right)/2$ as shown in Fig.3. Thus we have shown that $h_\alpha  \left( x \right)$ is always negative on the interior of domain $\mathcal{D}_1$ which indicate that $l_\alpha  \left( x \right)$ is a strictly monotone-decreasing function for $0 < x < 1, 1 < \alpha  < \left( {\sqrt {13}  - 1} \right)/2$. Similarly, we can show that $l_\alpha  \left( x \right)$ is a strictly monotone-increasing function for $\left( {\sqrt {7}  - 1} \right) < \alpha  < 1$. In this case, it is enough to prove the non-negative of the function $h_\alpha \left( {x } \right):={dl_\alpha  \left( x \right)/dx}$ on the domain $\mathcal{D}_2  = \left\{ {\left( {\alpha ,x} \right)|\left( {\sqrt {7}  - 1} \right)/2 \leq \alpha  \leq 1,0 \le x \le 1} \right\}$. Because $h_\alpha \left( {x } \right)$ has no critical points on the interior of $\mathcal{D}_2$ as shown in Fig.1, we consider the function value of $h_\alpha \left( {x } \right)$ on the boundary of $\mathcal{D}_2$. If $x\to1$, we can verify that the function $h_\alpha  \left( x \right)|_{x \to 1}$ is always positive for $\left( {\sqrt {7}  - 1} \right)/2 < \alpha  < 1$. If $\alpha = \left( {\sqrt {7}  - 1} \right)/2$, it is shown in Fig.4 that $h_\alpha  \left( x \right)|_{\alpha  = \left( {\sqrt {7}  - 1} \right)/2}$ is always positive for $0<x<1$. Therefore, $h_\alpha \left( {x } \right)$ is always positive for $\left( {\sqrt {7}  - 1} \right)/2 < \alpha  < 1$ which indicates that $l_\alpha \left( {x } \right)$ is a strictly monotone-increasing function in this case. Combining Eq.(\ref{q15}) we can derive $x_0=y_0$. However, from Eq.(\ref{q12})-(\ref{q14}) and $\nabla g_\alpha  \left( {x_0 ,x_0 } \right) = (0,0)$ we have $l_\alpha  \left( {\sqrt 2 x_0 } \right) = l_\alpha  \left( {x_0 } \right)$ which contradicts to the strict monotonicity of $l_\alpha  \left( x \right)$ for $0<x<1$. Therefore, we conclude that $g_\alpha  \left( {x,y} \right)$ has no critical points in the interior of $\mathcal{D}$.

\begin{figure}[ptb]
\includegraphics[scale=0.4,angle=0]{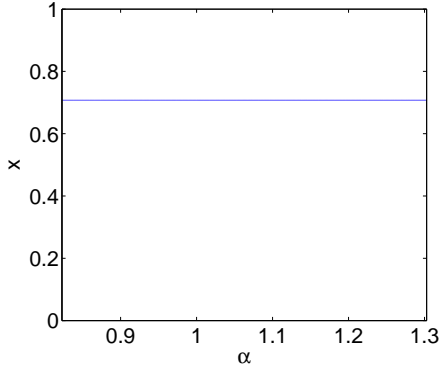}\caption{(color online). The plot of $\partial m_\alpha  \left( x \right)/\partial x =0$ for $0\leq x\leq1, \left( {\sqrt {7}  - 1} \right)/2 \leq \alpha \leq \left( {\sqrt {13}  - 1} \right)/2 $.}%
\label{fig5}
\end{figure}

Next, we consider the function value of $g_\alpha  \left( {x,y} \right)$ on the boundary of $\mathcal{D}$. If $x=0$ or $y=0$, it is direct to check that $g_\alpha  \left( {x,y} \right)=0$. When $x^2+y^2=1$, $g_\alpha  \left( {x,y} \right)$ becomes a two-variable function

\beqa\label{q18}
 m_\alpha  \left( x \right): = 1 - \frac{1}{{1 - a}}\log \left[ {\left( {\frac{{\Theta \left( x \right)}}{2}} \right)^a  + \left( {\frac{{\Xi \left( x \right)}}{2}} \right)^a } \right] \nonumber\\
  - \frac{1}{{1 - a}}\log \left[ {\left( {\frac{{\Theta \left( {\sqrt {1 - x^2 } } \right)}}{2}} \right)^a  + \left( {\frac{{\Xi \left( {\sqrt {1 - x^2 } } \right)}}{2}} \right)^a } \right].
\eeqa
As shown in Fig.5, $\partial m_\alpha  \left( x \right)/\partial x = 0$ has only one solution $x = 1/\sqrt 2$ on the domain $\mathcal{D}_3  = \left\{ {\left( {\alpha ,x} \right)|\left( {\sqrt {7}  - 1} \right)/2 \leq \alpha  \leq \left( {\sqrt {13}  - 1} \right)/2,0 \le x \le 1} \right\}$. On the other hand, we plot $\partial m_\alpha  \left( x \right)/\partial \alpha |_{x = 1/\sqrt 2 }$
in Fig.6 and we can see that the function is always positive for $\left( {\sqrt {7}  - 1} \right)/2 \leq \alpha  \leq \left( {\sqrt {13}  - 1} \right)/2$, which shows that $m_\alpha \left( x \right)$ has no critical points on the interior of domain $\mathcal{D}_3$. Then we consider the value of $m_\alpha  (x)$ on the boundary of $\mathcal{D}_3$. If $x=0$ or $1$, we have $m_\alpha  (x)=0$. When $\alpha=\left( {\sqrt {7}  - 1} \right)/2$ or $\alpha=\left( {\sqrt {13}  - 1} \right)/2$, it is direct to check that $m_\alpha  \left( x \right)$ is always a non-positive function. In Fig.7 we
plot $m_\alpha  \left( x \right)$ as a function of $x$ and $\alpha$, which illustrates our result.

\begin{figure}[ptb]
\includegraphics[scale=0.7,angle=0]{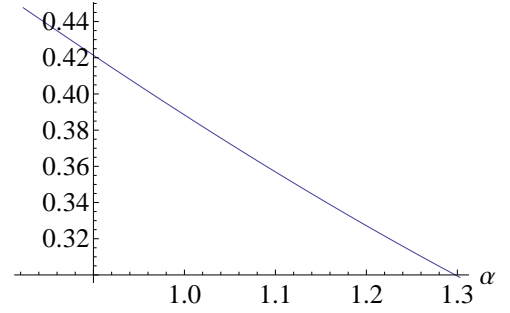}\caption{(color online). The plot of $\partial m_\alpha  \left( x \right)/\partial \alpha |_{x = 1/\sqrt 2 }$ as a function of $\alpha$ for $\left( {\sqrt {7}  - 1} \right)/2 \leq \alpha \leq \left( {\sqrt {13}  - 1} \right)/2 $.}%
\label{fig6}
\end{figure}

Combining the case for $\alpha=1$ which has been proved in Ref.\cite{Buscemi09pra}, we have completed the proof of Lemma 2. $\hfill\square$

\begin{figure}[ptb]
\includegraphics[scale=0.7,angle=0]{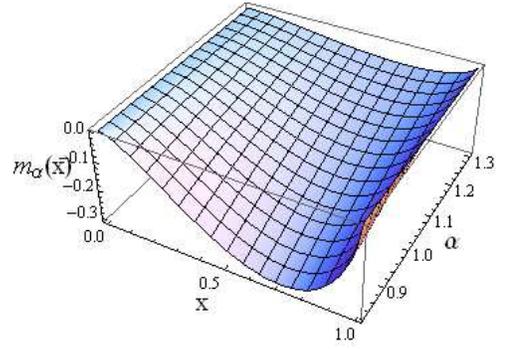}\caption{(color online). The plot of $m_\alpha  \left( x \right)$ as a function of $x$ and $\alpha$ for $0\leq x\leq1, \left( {\sqrt {7}  - 1} \right)/2 \leq \alpha \leq \left( {\sqrt {13}  - 1} \right)/2 $.}%
\label{fig7}
\end{figure}

Now we can prove the main result of this paper.

\emph{Theorem.} For $\left( {\sqrt {7}  - 1} \right)/2\leq \alpha \leq \left( {\sqrt {13}  - 1}\right)/2$, and any $n$-qubit state $\rho _{A_1 A_2  \cdots A_n }$, we have

\beqa\label{q19}
 {E_\alpha^a \left(\rho _{A_1 |A_2  \cdots A_n }\right)}    \le  {E_\alpha^a \left( {\rho _{A_1 A_2 } } \right)} +  \cdots  +  {E_\alpha^a \left( {\rho _{A_1 A_n } } \right)}, \nonumber\\
\eeqa
where ${E_\alpha^a \left(\rho _{A_1 |A_2  \cdots A_n }\right)} $ denotes the REoA in the partition $A_1|A_2\cdots A_n$, and ${E_\alpha^a \left( {\rho _{A_1 A_i } } \right)}$ is the REoA of the two-qubit subsystem $A_1A_i$ for $i=2,\ldots,n$.

\emph{Proof.} We first prove the polygamy relation for the pure state ${\left| \psi  \right\rangle _{A_1 |A_2  \cdots A_n } }$. Assuming that $ {C^2\left( {\rho _{A_1 |A_2  \cdots A_n } } \right)}   \le \left[ {C^a \left( {\rho _{A_1 A_2 } } \right)} \right]^2  +  \cdots  + \left[ {C^a \left( {\rho _{A_1 A_n } } \right)} \right]^2  \le 1$ in Eq.(\ref{q2}), then we have

\beqa\label{q20}
 &&{E_\alpha \left( {\left| \psi  \right\rangle _{A_1 |A_2  \cdots A_n } } \right)}    = f_\alpha \left( {C\left( {\rho _{A_1 |A_2  \cdots A_n } } \right)} \right) \nonumber\\
  &\le& f_\alpha  \left( {\sqrt {\left[ {C^a \left( {\rho _{A_1 A_2 } } \right)} \right]^2  +  \cdots  + \left[ {C^a \left( {\rho _{A_1 A_n } } \right)} \right]^2 } } \right) \nonumber\\
  &\le& f_\alpha  \left( {C^a \left( {\rho _{A_1 A_2 } } \right)} \right) \nonumber\\
  &+& f_\alpha  \left( {\sqrt {\left[ {C^a \left( {\rho _{A_1 A_3 } } \right)} \right]^2  +  \cdots  + \left[ {C^a \left( {\rho _{A_1 A_n } } \right)} \right]^2 } } \right) \nonumber\\
  &\le& f_\alpha  \left( {C^a \left( {\rho _{A_1 A_2 } } \right)} \right) +  \cdots  + f_\alpha  \left( {C^a \left( {\rho _{A_1 A_n } } \right)} \right) \nonumber\\
  &\le&  {E_\alpha^a \left( {\rho _{A_1 A_2 } } \right)}    +  \cdots  +  {E_\alpha^a \left( {\rho _{A_1 A_n } } \right)},
\eeqa
where in the first inequality we have used the monotonically increasing property of $f_\alpha(x)$ for $\alpha \ge \left( {\sqrt 7  - 1} \right)/2$, the second and third inequalities are obtained by the successive application of Lemma 2, and the last inequality is due to Lemma 1.

Then we consider the case $ {C^2\left( {\rho _{A_1 |A_2  \cdots A_n } } \right)}   \le 1 \le \left[ {C^a \left( {\rho _{A_1 A_2 } } \right)} \right]^2  +  \cdots  + \left[ {C^a \left( {\rho _{A_1 A_n } } \right)} \right]^2$. There must exist $k \in \left\{ {2, \ldots ,n - 1} \right\}$ such that  $\left[ {C^a \left( {\rho _{A_1 A_2 } } \right)} \right]^2  +  \cdots  + \left[ {C^a \left( {\rho _{A_1 A_k } } \right)} \right]^2  \le 1, \left[ {C^a \left( {\rho _{A_1 A_2 } } \right)} \right]^2  +  \cdots  + \left[ {C^a \left( {\rho _{A_1 A_{k + 1} } } \right)} \right]^2  > 1$. By defining $T: = \left[ {C^a \left( {\rho _{A_1 A_2 } } \right)} \right]^2  +  \cdots  + \left[ {C^a \left( {\rho _{A_1 A_{k + 1} } } \right)} \right]^2  - 1 > 0$, we can derive

\beqa\label{q21}
&&  {E_\alpha \left( {\left| \psi  \right\rangle _{A_1 |A_2  \cdots A_n } } \right)}    = f_\alpha  \left( {C\left( {\rho _{A_1 |A_2  \cdots A_n } } \right)} \right) \le f_\alpha  \left( 1 \right) \nonumber\\
  &=& f_\alpha \left( {\sqrt {\left[ {C^a \left( {\rho _{A_1 A_2 } } \right)} \right]^2  +  \cdots  + \left[ {C^a \left( {\rho _{A_1 A_{k + 1} } } \right)} \right]^2  - T} } \right) \nonumber\\
  &\le& f_\alpha  \left( {\sqrt {\left[ {C^a \left( {\rho _{A_1 A_2 } } \right)} \right]^2  +  \cdots  + \left[ {C^a \left( {\rho _{A_1 A_k } } \right)} \right]^2 } } \right) \nonumber\\
  &+& f_\alpha  \left( {\sqrt {\left[ {C^a \left( {\rho _{A_1 A_{k + 1} } } \right)} \right]^2  - T} } \right) \nonumber\\
  &\le& f_\alpha  \left( {C^a \left( {\rho _{A_1 A_2 } } \right)} \right) +  \cdots  + f_\alpha  \left( {C^a \left( {\rho _{A_1 A_{k + 1} } } \right)} \right) \nonumber\\
  &\le&  {E_\alpha^a \left( {\rho _{A_1 A_2 } } \right)}    +  \cdots  +  {E_\alpha^a \left( {\rho _{A_1 A_n } } \right)},
\eeqa
where we have used the monotonically increasing property of $f_\alpha(x)$ in the first equality, in the second inequality we have used Lemma 2, the third inequality is obtained by the successive application of Lemma 2, and the last inequality is due to Lemma 1.

Using the polygamy relation for the pure state we can prove the Theorem in the mixed state. Suppose that the optimal decomposition for $E_\alpha ^a \left( {\rho _{A_1 |A_2  \cdots A_n } } \right)$ is $\left\{ {p_j ,\left| {\psi _j } \right\rangle _{A_1 |A_2  \cdots A_n } } \right\}$, then we have

\beqa\label{q22}
 &&E_\alpha ^a \left( {\rho _{A_1 |A_2  \cdots A_n } } \right) = \sum\nolimits_i {p_j } E_\alpha  \left( {\left| {\psi _j } \right\rangle _{A_1 |A_2  \cdots A_n } } \right) \nonumber\\
  &\le& \sum\nolimits_j {p_j } \left( {E_\alpha ^a \left( {\rho _{A_1 A_2 }^j } \right) +  \cdots  + E_\alpha ^a \left( {\rho _{A_1 A_n }^j } \right)} \right) \nonumber\\
  &\le& E_\alpha ^a \left( {\rho _{A_1 A_2 } } \right) +  \cdots  + E_\alpha ^a \left( {\rho _{A_1 A_n } } \right),
\eeqa
where ${\rho _{A_1 A_j }^j }$ is the reduced density matrix of ${\left| {\psi _j } \right\rangle _{A_1 A_2  \cdots A_n } }$ onto the two-qubit subsystem $A_1 A_j$ for each $i=2,\ldots,n$, in the first inequality we have used the polygamy relation for each pure state decomposition state ${\left| {\psi _j } \right\rangle _{A_1 |A_2  \cdots A_n } }$
\beqa\label{q23}
E_\alpha  \left( {\left| {\psi _j } \right\rangle _{A_1 |A_2  \cdots A_n } } \right) \le E_\alpha ^a \left( {\rho _{A_1 A_2 }^j } \right) +  \cdots  + E_\alpha ^a \left( {\rho _{A_1 A_n }^j } \right),\nonumber\\
\eeqa
and the last inequality is due to the definition of REoA for each ${\rho _{A_1 A_i } }$. Thus we have completed the proof of Theorem. $\hfill\square$

Furthermore, we can establish the following $\mu$th power polygamy inequalities for the R\'{e}nyi-$\alpha$ entanglement of assistance.

\emph{Corollary.} For $\left( {\sqrt {7}  - 1} \right)/2\leq \alpha \leq \left( {\sqrt {13}  - 1}\right)/2, 0\leq\mu\leq1$, and any $n$-qubit state $\rho _{A_1 A_2  \cdots A_n }$, we have
\beqa\label{q24}
\left[ {E_\alpha ^a \left( {\rho _{A_1 |A_2  \cdots A_n } } \right)} \right]^\mu   \le \left[ {E_\alpha ^a \left( {\rho _{A_1 A_2 } } \right)} \right]^\mu   +  \cdots  + \left[ {E_\alpha ^a \left( {\rho _{A_1 A_n } } \right)} \right]^\mu \nonumber\\.
\eeqa
This inequality holds because $\left[ {E_\alpha ^a \left( {\rho _{A_1 |A_2  \cdots A_n } } \right)} \right]^\mu
  \le \left[ {E_\alpha ^a \left( {\rho _{A_1 A_2 } } \right) +  \cdots  + E_\alpha ^a \left( {\rho _{A_1 A_n } } \right)} \right]^\mu
  \le \left[ {E_\alpha ^a \left( {\rho _{A_1 A_2 } } \right)} \right]^\mu   +  \cdots  + \left[ {E_\alpha ^a \left( {\rho _{A_1 A_n } } \right)} \right]^\mu$,
where the last inequality is due to the concave property of $x^\mu$ for $0\leq\mu\leq1$.

By introducing the dual concept of REoA, we have established polygamy relations for the R\'{e}nyi-$\alpha$ entanglement of assistance in multi-qubit systems. We have also generalized the polygamy inequalities into the $\mu$th power of REoA. These derived polygamy relations provide a lower bound for distribution of bipartite REoA in a multi-party system. The monogamy and polygamy relations are not only fundamental property of entanglement in multi-party systems but also provide us an efficient way of characterizing multipartite entanglement. In Ref. \cite{Song16pra}, we have proved that squared R\'{e}nyi-$\alpha$ entanglement with the order $\alpha\geq (\sqrt{7}-1)/2$ obeys a general monogamy relation in an arbitrary $n$-qubit mixed state. It is further shown that we can construct the multipartite entanglement indicators in terms of ER$\alpha$E which still work well even when the indicators based on the concurrence and EoF lose their efficacy. Thus our polygamy inequalities together with previous monogamy inequalities in terms of ER$\alpha$E might provide a useful tool to understand the property of multi-party quantum entanglement.

This work was supported by NSF-China under Grant Nos.11374085, 11274010, the Anhui Provincial Natural Science Foundation under Grant Nos.1708085MA12, 1708085MA10, the Key Program of the Education Department of Anhui Province under Grant Nos. KJ2017A922, KJ2016A583, the discipline top-notch talents Foundation of Anhui Provincial Universities under Grant Nos.gxbjZD2017024, gxbjZD2016078, the Anhui Provincial Candidates for academic and technical leaders Foundation under Grant No.2015H052 and the Excellent Young Talents Support Plan of Anhui Provincial Universities.

\end{document}